\documentclass[preprint]{aastex631}
\shorttitle{Witt's Hyperbola: a Constraint on Lens Models}
\graphicspath{{./}{figures/}}

\begin{document}


\title{
  Witt's hyperbola is both predicted and observed to pass close to the lensing
  galaxies in quadruple quasars
}

\correspondingauthor{Paul L. Schechter}
\email{schech@mit.edu}

\author[0000-0002-5665-4172]{Paul L. Schechter}
\affiliation{MIT Kavli Institute and Department of Physics \& MIT Kavli Institute McNair Building 37-635, 77 Massachusetts Ave, Cambridge, MA, 02139,USA}

\author[0000-0003-2187-7090]{Richard Luhtaru}
\affiliation{MIT Department of Physics   Cambridge, MA 02139, USA}










\begin{abstract}

When a rectangular hyperbola is constructed from the image positions
of a quadruply lensed quasar, as proposed by Witt (1997), it passes
very close to the the lensing galaxy.  The median measured
perpendicular offset between the observed light center of the lens and
Witt's hyperbola is $0\farcs013$ for a sample for 39 systems lensed by
a relatively isolated galaxy.  The family of lens models adopted by
Witt predicts that the lens lies on the hyperbola, but its position in
not used for its construction.  The median offset corresponds to
roughly 1\% of the Einstein ring radius, and suggests that the centers
of the lensing potential are close to the light centers of the lens.
By putting a restrictive prior on the perpendicular distance to Witt's
hyperbola (or on the distance between the galaxy and the potential),
one reduces by one the dimensionality of a model space when fitting
data.  Taking the brightest pixel of a lensing galaxy as its center
avoids a shortcoming of using the average light center for a
constraint.
  
\end{abstract}



\section{Eliminate free parameters when you can}\label{sec:eliminate}

Any model for a quadruply lensed quasar (or supernova) includes both a
model for the lensed object (a pointlike quasar, and sometimes an extended
host) and a model for the gravitational potential that deflects light
from the source.  However small the number of parameters associated
with any particular model, one would almost always choose to eliminate
a free parameter if it were possible without introducing an unmanageable
systematic error.

We argue here that, for the case of relatively isolated lensing
galaxies, the observed position of the lensing galaxy can be used to
reduce, in effect, the number of model parameters by one.  There are
roughly a half dozen independent codes used to model quadruply lensed
quasars.  Most of them use the position of the lensing galaxy to
constrain the center of the lensing potential, albeit with differences
in implementation.  The stringency adopted for these constraints varies
from one modeler to the next, with little or no discussion of the
choice.  And with a few exceptions exceptions e.g.  \citet{chen} there
is likewise no reporting of the derived offset of the lens potential
from the lensing galaxy.

Here we urge constraining the center of the lens potential using the
observed distance of the lensing galaxy from the rectangular hyperbola
constructed by \citet{witt}, using {\it only} the positions of the
four lensed quasar images and {\it not} that of the lensing galaxy.
While Witt's recipe is strictly geometric,
it was derived from family of physical models that {\it do} put the
center of the lensing potential on his hyperbola.  The observed
proximity of relatively isolated lenses to Witt's hyperbola
strongly suggests that the the centers of the lensing potentials
are likewise close to the hyperbola.

In Section 2 below we review key aspects of Witt's hyperbola.  In
Section 3 we report our principal result:
for a sample of 39 relatively isolated lensing
galaxies, the distances of their observed centers from Witt's hyperbola
are small compared to the systems' Einstein ring radii.
This suggest a prior for use in modeling similar systems.
In Section 4 we interpret our measured distances, contending that the family of
models that motivated Witt is a good first approximation for systems
with a single dominant lens.  In an Appendix, we advocate using
the brightest point at the center of the lens in calculating offsets
from Witt's hyperbola.

\begin{figure*}[t]
\includegraphics[width=1.00\textwidth]{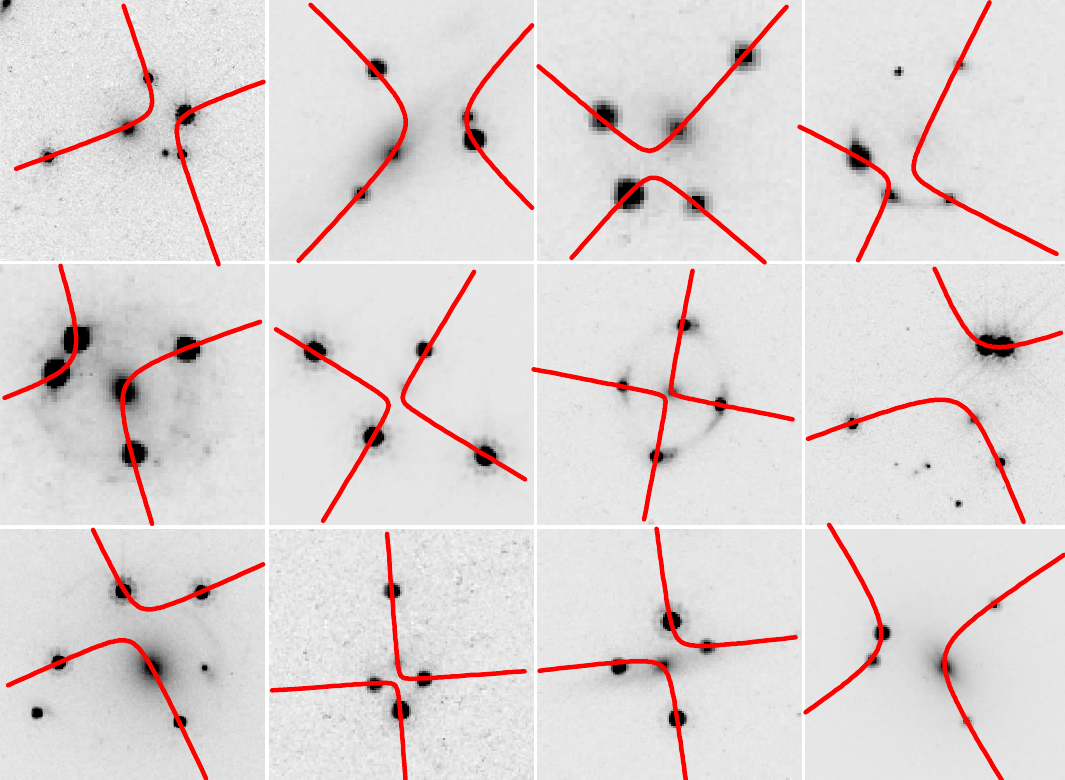}  
\caption{Witt's hyperbola constructed from the four quasar image
  positions for a dozen systems 
  lensed by galaxies that, in all but one case, are relatively isolated.
  They were observed with the F814W filter in an HST cycle 26 proposal
  and are shown superposed on those exposures. 
  The positions of the lensing galaxies were 
  {\it not} used in constructing the hyperbolae.
  The long, ``primary'' branch of the Witt hyperbola
  nonetheless passes very close to the lensing galaxy in every case,
  as  predicted by the family of models that motivated its construction.
  Left to right and top to bottom are the 
  systems J0659+16, J1330+18, J2205$-$37, J1042+16, J1131$-$44,
  J1134$-$21, J1537$-$30, J0818$-$26, J1721+88, J0029$-$38, J1817+27,
  and J2100$-$44}
\label{fig:hypsover12}
\end{figure*}

\section{Witt's geometrically constructed hyperbola}

Witt (1997) showed that {\it any} gravitational lens whose
equipotentials are concentric similar ellipses (characterized here
by their semi-ellipticity $\eta$) produces four images that lie on a
rectangular hyperbola, one with perpendicular asymptotes. As a
special case, Witt's hyperbola passes through the images produced
by singular isothermal elliptical potential (hereafter SIEP).

In addition, Witt showed that a singular isothermal circular
potential with external shear, $\gamma_{ext}$, (hereafter SIS+XS),
{\it also} produces four images that lie on a rectangular hyperbola.

Moreover there is a one dimensional family of combinations of the
SIS+XS and SIEP with shear parallel to ellipticity (henceforth
SIEP+XS$_{||}$ for all of whose members Witt's hyperbola passes through
the same quartet of observed quasar image positions, contingent on
$\gamma_{ext} + \eta$ being kept constant at some value governed by those
$\gamma_{ext} + \eta$ being kept constant at some value governed by thosefour positions.

While all of the members of a family with the same value of
$\gamma_{ext} + \eta$ produce the same four quasar images, the
position of the center of the potential (which also lies on the
hyperbola) varies with the relative importance of $\gamma_{ext}$ and
$\eta$.  This is extensively delineated by \citet{luhtaru}.

Though Witt's hyperbola was motivated by the SIEP and SIS+XS models,
the construction of Witt's hyperbola depends {\it only} on the
positions of the four quasar images.  It is a geometric construct,
and with hindsight might have been created without reference
to either of those models.  

Figure 1 shows Witt hyperbolae overlayed on HST F814W exposures of a
dozen lens systems, in all but one of which the lensing galaxy is
relatively isolated.  The positions of the four quasar images have
been used in Witt's geometric construction but the position of the
lensing galaxy has {\it NOT}.  The lensing galaxy nonetheless lies
very close to the ``primary'' branch of the Witt hyperbola in every
case.  In the next section we quantify this qualitative observation.

\section{Observed perpendicular distances of relatively isolated lens galaxies
from Witt's hyperbola
}
\subsection{What makes quadruply lensed quasars quadruple?}\label{subsec:WMQLQQ}

\citet{luhtaru} measured perpendicular distances of 39 relatively
isolated lens galaxies to Witt's hyperbola.   Those measurements
were a byproduct of their their primary objective, to
answer the question asked in the title of their paper,
``What Makes Quadruply Lensed Quasars Quadruple?'' 
They adopted the SIEP + XS$_{||}$ family of models introduced
in the previous section to determine the relative contributions
of shear and semi-ellipticity to the lensing potentials.  

In Appendix A we reanalyze 10 of the 39 systems.  The incidental
perpendicular distances reported there are the focus of the results
of the next subsection.  For the sake of completeness, we
review here some details of the  SIEP + XS$_{||}$ family of models.

The dimensionless projected potential, $\psi$ of the  SIEP + XS$_{||}$
model is given by
\begin{equation}
    \psi(b,\eta, \gamma_{ext}) = b\sqrt{q_{pot}x^2 + \frac{y^2}{q_{pot}}} - \frac{\gamma_{ext}}{2} (x^2-y^2)\, .
\end{equation}
where  $b$ is the
strength of the lens, and $\gamma_xext$ is
the shear.  \citet{luhtaru} define
a ``semi-ellipticity'', $\eta$, such that
\begin{equation}
  q_{pot} = (1-\eta)/(1+\eta) \quad ,
\end{equation}
and an ``effective quadrupole'', $\Gamma_{eff}$,
\begin{equation}
    \Gamma_{eff} \equiv \frac{\gamma_{ext} + \eta}{1 + \gamma_{ext} + \eta }  \quad ,
\end{equation}
in which  $\gamma_{ext}$ and $\eta$ are interchangeable.
For semi-ellipticities and shears typical of quadruply lensed quasars
\begin{equation}
  \Gamma_{eff} \approx \eta + \gamma_{ext}\quad.
\end{equation}    

The effective quadrupole, $\Gamma_{eff}$, is tightly constrained by
the elongation of the image configuration. The strength parameter $b$,
roughly equal to the Einstein radius, $\theta_E$, is also well
constrained by the four quasar images.  By contrast, the ratio of the
semi-ellipticity $\eta$ to the shear, $\gamma_{ext}$ is unconstrained,
by construction,

That ratio {\it can}, however, be estimated by drawing a line from
lensing galaxy perpendicular to Witt's hyperbola.  The position along
Witt's hyperbola determines the ratio.  The perpendicular distance,
$D_\perp$, is a byproduct that is central to the present paper. 

\subsection{Observed perpendicular distances from Witt's hyperbola} \label{subsec:observedDPERP}

The data  presented in Table 2 of \citet{luhtaru} were used to determine
perpendicular distances for the galaxies in their sample with
the exception of the 10 galaxies considered in \ref{app:newmeas}
for which our new measurements have been used.
In Table \ref{tab:39perpsONLY} we give perpendicular distances from the
lens to Witt's hyperbola for these systems.

\startlongtable
\begin{deluxetable}{lDlDlDlD}
\centerwidetable
\tabletypesize{\scriptsize}
\tablecaption{Observed distance of lensed galaxy from Witt's hyperbola for 39 single-lens systems} \label{tab:39perpsONLY}
\tablehead{
 \colhead{System name} & \multicolumn{2}{c}{$D_\perp(\arcsec)$} & \colhead{System name} & \multicolumn{2}{c}{$D_\perp(\arcsec)$} & \colhead{System name} & \multicolumn{2}{c}{$D_\perp(\arcsec)$} & \colhead{System name} & \multicolumn{2}{c}{$D_\perp(\arcsec)$}  \\
}
\decimals
\startdata
J0029-3814       & 0.011  & B0712+472        & 0.001  & SDSS J1251+2935  & 0.005 & J1721+8842       & 0.016  \\
J0030-1525       & 0.002  & HS0810+2554      & 0.015  & HST12531-2914    & 0.009 & J1817+27         & 0.051  \\
PS J0147+4630    & 0.063  & RXJ0911+0551     & 0.001  & SDSS J1330+1810  & 0.005 & WFI2026-4536     & 0.013  \\ 
SDSS J0248+1913  & 0.005  & SDSS0924+0219    & 0.000  & HST14113+5211    & 0.022 & DES J2038-4008   & 0.006  \\ 
ATLAS J0259-1635 & 0.020  & J1042+1641       & 0.035  & H1413+117        & 0.033 & B2045+265        & 0.014  \\
DES J0405-3308   & 0.014  & PG1115+080       & 0.007  & HST14176+5226    & 0.075 & J2100-4452       & 0.014  \\   
DES J0420-4037   & 0.002  & RXJ1131-1231     & 0.036  & B1422+231        & 0.007 & J2145+6345       & 0.062  \\    
HE0435-1223      & 0.008  & J1131-4419       & 0.004  & SDSS J1433+6007  & 0.071 & J2205-3727       & 0.000  \\  
J0530-3730       & 0.017  & J1134-2103       & 0.042  & J1537-3010       & 0.004 & WISE J2344-3056  & 0.001  \\  
J0659+1629       & 0.077  & SDSS1138+0314    & 0.004  & PS J1606-2333    & 0.033                    &        \\
\enddata
\end{deluxetable}

\begin{figure*}[t]\centering
\includegraphics[width=0.5\textwidth]{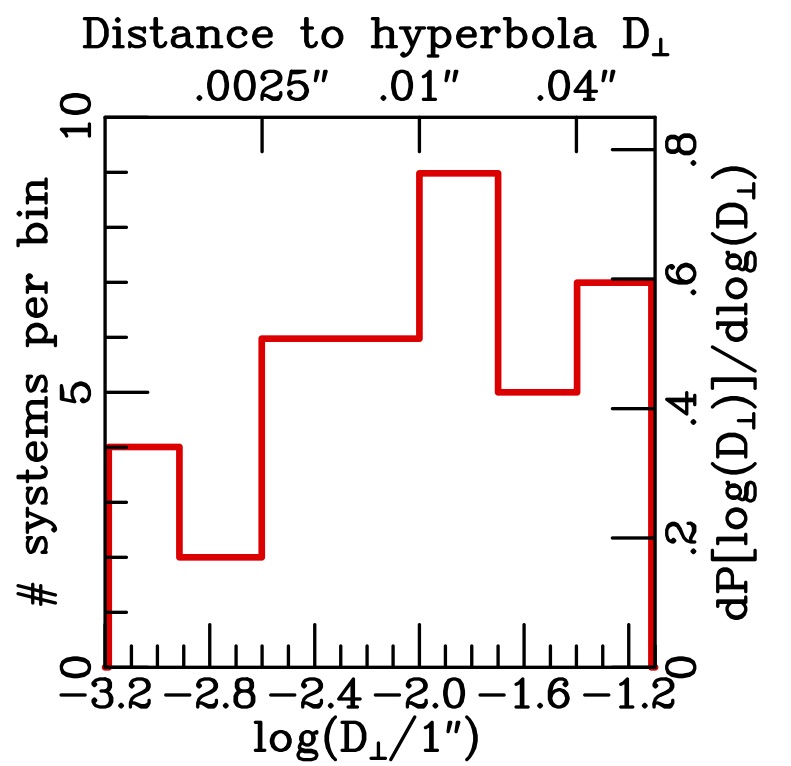}
\caption{The distribution of perpendicular distances, $D_\perp$, to
  Witt's hyperbola measured for the \citet{luhtaru} sample of 39
  quasars quadruply lensed by relatively isolated galaxies.  The scale
  on the left gives the number of systems in each bin.  The bins have
  constant logarithmic width $dlog(D_\perp/1\arcsec) = log 2$, so that
  each bin spans a factor of two, as shown by the top scale.  The
  scale on the right gives the differential probability
  $dP[\log(D_\perp)]$ of $\log(D_\perp)$ per logarithmic increment
  and integrates to unity over the range shown.}
\label{fig:luhthist}
\end{figure*}

The results of these measurements are presented as a histogram in Figure
\ref{fig:luhthist}. There are two sets of labels for each axis.
The upper axis gives the perpendicular distance, $D_\perp$ in
arcseconds.  The lower axis gives the decimal logarithm of
$D_\perp$.  The  7 bins in the histogram are logarithmic,
with $\Delta D_\perp = 0.301$, corresponding to a factor of two.
The left hand axis gives the number of systems in each bin.

As discussed in Appendix \ref{app:newmeas}, we estimate the errors in our
measured galaxy positions to be {\it roughly} $0\farcs005$.  Those
erros dominate the $D_{perp}$ measurements on the left half of the
histogram.  But their contribution to the $D_{perp}$ measurements on
the right half are likely to be small.  We take the $D_\perp$
measurements on the right half of the histogram to reflect
shortcomings in the SIEP+XS$_{||}$ model.
  
The right hand axis of Figure \ref{fig:luhthist} gives the differential
probability $dP$ of finding a member of the sample per unit common
logarithm, $d \log (D_\perp)$.  One obtains it by dividing the number
of systems observed in a bin by the sample size, 39, and scaling by
$1/\log 2$, since we have used bins that span only a factor of two in
$D_\perp$.  Using this axis the red curve integrates to unity.  Given
the small numbers, one might with equanimity summarize  the
result of Figure \ref{fig:luhthist} as $dP[log (D_\perp)]/d \log
(D_\perp) = 1/2$ over the range plotted..

It should be noted that average lens strength $b$ for the
seven systems in the rightmost bin is $<b> = 1\farcs5$.
One would have a yet tighter constraint if one put a prior
on $D_\perp / b$.

It should also be remembered that the leftmost bins are dominated by
errors in the measurement of galaxy centers, which are likely to be
different for different data sets.

\section{A prior on the center of the lensing potential}

The median $D_\perp$ in Figure \ref{fig:luhthist} is $0\farcs013$.  The
median value of the lens strength, $b$, is $0\farcs85$.  This puts the
typical lensing galaxy at a distance from Witt's hyperbola that is
1/70th radius of the Einstein ring.

\subsection{What explains the proximity of Witt's hyperbola
  to the lensing galaxy?}

There are two categories of explanations for the proximity
of the lensing galaxy to Witt's hyperbola in Figure \ref{fig:luhthist}.
The first of these posits that the SIEP+XS$_{||}$ is
esssentially correct and that the center of the potential
does in fact lie on Witt's hyperbola.  In this case the
lensing galaxy is offset from the center of the potential,
much as the centers of brightest cluster galaxies are
offset from the the centers of their potentials \citep{vandenbosch}.

The second category assumes that the potential {\it is} centered on
the lensing galaxy, but that it is is more complicated than
SIEP+XS$_{||}$ model.  The positions of the four quasar images are
therefore displaced from the SIEP+XS$_{||}$ predictions and a
hyperbola that is consequently displace from the center of the
potential.  Examples would be equipotentials for which either the
ellipticity or their orientation changes as a function of radius.

\subsection{Constraining the center of the lensing potential}

For the first category of explanations, the distribution of offsets show in
Figure \ref{fig:luhthist} is an appropriate prior on the center of the
potential for use in modeling quads.

For the second category of explanations, the appropriate prior on the
offset of the center of the potential from the center of the lens is a
delta function.  On the assumption that both explanations are at work,
the prior derived from Figure \ref{fig:luhthist} is {\it conservative}.  We
recommend its use in modeling quadruply lensed quasars.

Taking the explanation of the offsets of the lens of the potential to
arise entirely from unmodeled complexities in the true potential
(category 1), modelers can still use the perpendicular distances in
\ref{fig:luhthist} even if they are disinclined to adding this term in
the quantity they minimize.  One can assume that those complexities
introduce an offset of the hyperbola from point where the
perpendicular intersects in the direction orthogonal to the
perpendicular that is equal to the perpendicular offset.  Multiplying
the abscissas in Figure \ref{fig:luhthist} by $\sqrt{2}$, the figure can be
taken as a prior on the offset of the center of the potential from the
center of light.

\section{Conclusions}\label{sec:conclusions}

We have shown when a relatively isolated galaxy produces a quadruply lensed
quasar, its center lies close to the hyperbola constructed by
\citet{witt} from the postions of the four quasar images.
This permits the use of a strong prior the position of the center
of the lensing potential. in effect reducing by one the number of parameters
model parameters used to model such a system.

\begin{acknowledgments}
  We thank Sebastian Ertl for communicating results in advance of publication.
  We also thank an anonymous referee for a careful reading that led to
  a complete overhaul of the paper, from title to conclusions.
\end{acknowledgments}

%





\vfill
\eject
\appendix
\section{New measurements of the centers of galaxies
  quadruply lensing quasars}\label{app:newmeas}

\subsection{Why does the astrometric accuracy of lens positions matter?}\label{subsec:doBETTER}

We did not set out to study the proximity of lensing galaxies
to Witt's hyperbola reported in Section \ref{subsec:observedDPERP}.
That finding emerged from an effort to improve upon the estimate of the relative
contributions of the flattening of lens galaxies on one hand and tidal shear
on the other to the
quadrupole moments of potentials obtained by \citet{luhtaru}, whose approach
is described  in Section \ref{subsec:WMQLQQ}.

Their scheme is limited by the accuracy with which one can measure the
positions of the quasar images and the position of the lensing galaxy,
which is assumed to have a singular isothermal elliptical potential
with parallel external shear (the SIEP+XS$_{||}$ model).
The uncertainty in the flattening/shear decomposition is estimated
from the offset the lensing galaxy from the \citet{witt} hyperbola.

Luhtaru et al.\ concluded that shear contributes roughly twice
as much as galaxy flattening to the ``effective shear'', $\Gamma_{eff}$,
the overall quadrupole component of a potential that quadruply lenses
a quasar. Were the uncertainties in their adopted positions larger than
reported, the reliability of that conclusion might be challenged.

Measuring positions for the four images of a quadruply lensed quasar is
less straightforward than one might at first imagine.  The quasars are
embedded in host galaxies and their images are projected onto the
lensing galaxies.  Perfect modeling of an exposure must take
into account the point spread function, the various shapes and sizes
of galaxies (lenses and quasar hosts), and the gravitational potential
that produces four images from a single quasar and host.  While multiple
software suites are widely available to analyze quadruply lensed
quasar systems, each of them incorporates different choices in
accounting for the relative importance of these features.

\subsection{Discrepant reports of image and lens center positions}

\cite{ertl} compared positions for the images
of nine quadruply lensed quasars measured using two different
automated schemes on the same set of HST images.  One of those was a
version of {\tt Lenstronomy} developed by \cite{schmidt}
and described therein.  The other was a version of
{\tt GLEE}, developed and described by \citet{ertl}

After discarding one extremely discrepant system, 
in the quasar image positions measured using the two schemes
differed by $0\farcs006$ and
$0\farcs005$ in right ascension and declination.  This was larger than
had been anticipated.  Comparing the {\tt GLEE} results with
astrometry from Gaia DR3 \citep{brown}, Ertl et al.\  reported
differences of $0\farcs0017$ and $0\farcs0023$, respectively.

\citet{luhtaru} was based on astrometry drawn from the literature, much of it
from the CASTLES \citep{castles} data base.  They also reported new
measurements on thirteen recently acquired HST exposures.  They
described their measurements, using a program called {\tt clumpfit}, as
``preliminary'' deferring to the anticipated measurements by \citet{schmidt}
which they called ``authoritative."  When Ertl et al. compared their {\tt GLEE}
astrometry with the ``preliminary'' results of \citet{luhtaru},
they found rms differences marginally smaller than with Gaia DR3, $0\farcs0017$ in
each direction.  It would therefore seem that Luhtaru et al.\ were mistaken in
deferring to the \citet{schmidt} measurements. 

Luhtaru et al.\ originally used measurements for ten systems reported
by \cite{shajib} that were obtained with an earlier version of {\tt
  lenstronomy}.\footnote{\citet{luhtaru} chose to use astrometry for
  an eleventh system, SDSS1310+18, from \citet{rusu} rather
  than that of \citet{shajib}.}  If the errors in these were as large
as those in the \citet{schmidt} astrometry, the Luhtaru et al.\
conclusion regarding the relative importance of shear and ellipticity
might be compromised.  We therefore remeasured, using {\tt clumpfit}, the same ten HST images
analyzed by \citet{shajib}.

In comparing {\tt clumpfit} and {\tt GLEE}, we were impressed that
the agreement for some systems was better than $0\farcs001$.  With
positions this good, the uncertainties in the modeled centers
of the potentials would be correspondingly accurate.  This motivated
us to compare the {\tt clumpfit} positions for the maximum in lens surface
brightness with the theoretical centers for the potential computed
from the positions of the quasar images.

\subsection{{New \tt clumpfit} astrometry}

The \citet{luhtaru} measurements were carried out with the program
{\tt clumpfit} described briefly in \citet{Agnello}.  It incorporates
elements of {\tt DoPHOT} \citep{Schechter93}, fitting point sources
with either an elliptical ``pseudo-Gaussian'' or a lookup table taken
from a nearby star.  Galaxies are fit with elliptical
pseudo-Gaussians.  These do not allow for curved images and so are ill
suited toward modeling the images of quasar hosts.  The {\tt clumpfit}
model does not couple the images to a lensing potential, which can be
modeled subsequently using the image parameters.

We used nearby stars as PSF templates when possible.  Unfortunately,
for the majority of systems there were no stars on the HST images
sufficiently bright for this purpose, so pseudo-Gaussians were used.
For some of the systems for which PSF templates were available we
also tried pseudo-Gaussians and found no substantial
differences.

The position of the lensing galaxy is of primary importance in
distinguishing between ellipticity and shear, so the {\tt clumpfit}
residuals were examined for effects that might influence the position
of the lensing galaxy.

The two greatest concerns were that a) the very substantial residuals from the
four point sources would ``pull'' or ``push'' the outer isophotes of the much
lower surface brightness lensing galaxy and b) that the unmodeled flux
from the quasar host might similarly pull those isophotes.

To minimize this effect, the lensing galaxy was often taken to be
smaller than its isophotes would indicate, so as to weight the light
from the center of the lens more heavily.  In some cases, the width of
the pseudo-Gaussian model was fixed at a value smaller than the best
fitting value.  In others, the lens galaxy was taken to have the PSF of
a stellar template.  Either way, the central intensity of the model
for the starlight was adjusted by hand to exactly account for the flux
at the center of the lens.

\begin{figure*}[t]
  \hspace*{1.35cm}\includegraphics[scale=0.30]{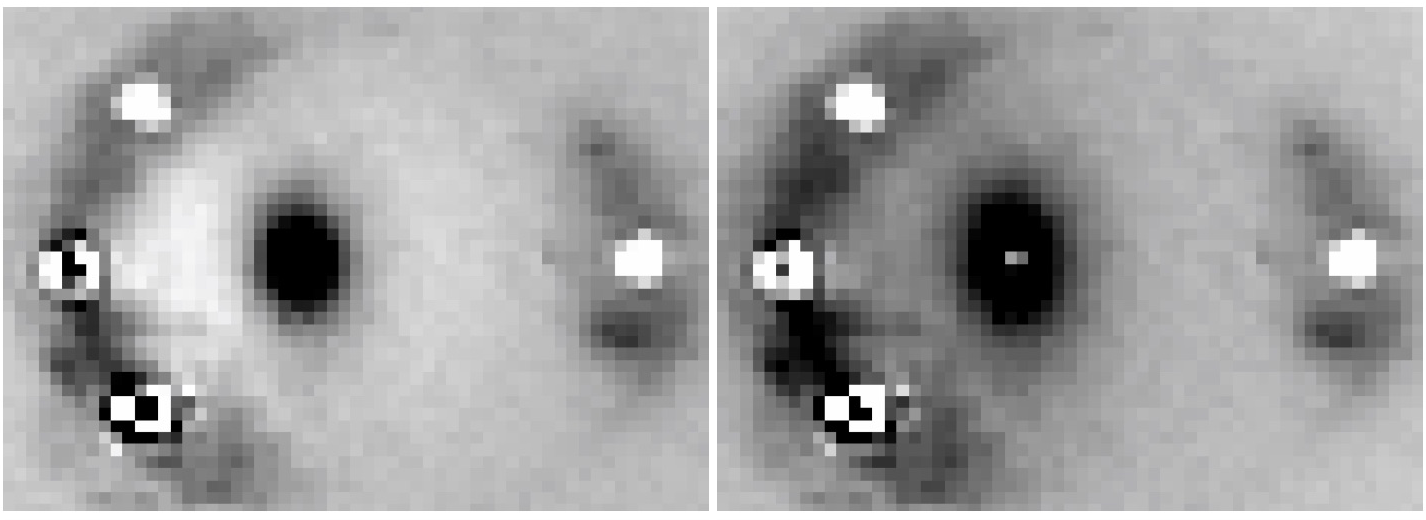}
  \caption{The left and right panels show residuals at the same contrast
    from fits of two different models to an HST F814W exposure of
    J1251+29.  In the left panel, the lensing galaxy was
    taken to be an elliptical
    pseudo-Gaussian.  In the right panel, it was taken to be a point
    source producing zero residual flux in the central pixel.
    The residuals in the left panel are asymmetric, the result
    of the outer isophotes of the elliptical Gaussian heing 
    pulled toward the extended structure of the quasar host. 
    }
\label{fig:extendedVpsf}
\end{figure*}

The effect is illustrated in Figure \ref{fig:extendedVpsf}, which shows
results from two alternative fits of surface brightness profiles to
an HST F814 image of the lensed quasar system J1251+29.  The 
left panel shows residuals when stellar PSFs are fit to the four images and
the central galaxy is taken to be a unconstrained elliptical
pseudo-Gaussian.  Note that the galaxy has been pulled toward the
extended structure of the quasar host, leaving positive residuals at
the center of the lensing galaxy and negative residuals toward the
triplet of quasar images.  Using this fit, \citet{luhtaru} found the
system to be shear-dominated.

The right panel shows residuals from the fit of the stellar PSF to
the center of the galaxy, with the amplitude constrained to give zero
residual in the brightest pixel.  The resulting center is our best
estimate of the true position,  giving shear and ellipticity
contributing equally.  This result is still somewhat suspect, as the
angle of the resultant quadrupole lies very close to the position
angle of the isophotes of the lensing galaxy.

The lesson to be learned (or reinforced) from Figure \ref{fig:extendedVpsf} is
that residuals tell you more than side-by-side comparisons of
models and data.\footnote{Secondarily, white sky (inverse grayscale) representations do better at bringing out low surface brightness residuals.}

As we worked through the sample, system by system, to improve upon the derived
centroids, different schemes were developed on an ad hoc basis.
The effects of imperfect point source subtraction on the lens position
could in some cases be minimized by measuring an F475X exposure rather
than the F814W exposure, but we this was done only when the F814W
frame gave a suspect result. For some systems the lensing
galaxy was too faint to make the F475X exposure useful.  Each of the
ten systems for which we present results was therefore treated
slightly differently, depending upon the particular circumstances.

 \begin{deluxetable} {lDDDDDDDD D             D              }
  
\centerwidetable
\tabletypesize{\scriptsize}
\tablecaption{Newly measured relative positions for galaxies and images of quadruply lensed quasars}\label{tab:newmeas}
\tablehead{
  \colhead{} &
  \colhead{} &  \colhead{} &  \colhead{} &  \colhead{} &
  \colhead{} &  \colhead{} &  \colhead{} &  \colhead{} &
  \colhead{} &  \colhead{} &  \colhead{} &  \colhead{} &
  \colhead{} &  \colhead{} &  \colhead{} &  \colhead{} \\  
%
  \colhead{} & \multicolumn{4}{c}{Image A} & \multicolumn{4}{c}{Image B} & \multicolumn{4}{c}{Image C} & \multicolumn{4}{c}{Image D} & \multicolumn{4}{c}{Galaxy}               \\  
%
 \colhead{System name} & \twocolhead{$\Delta \alpha \cos \delta$} & \twocolhead{$\Delta \delta$} & \twocolhead{$\Delta \alpha \cos \delta$} & \twocolhead{$\Delta \delta$} & \twocolhead{$\Delta \alpha \cos \delta$} & \twocolhead{$\Delta \delta$} & \twocolhead{$\Delta \alpha \cos \delta$} & \twocolhead{$\Delta \delta$} & \multicolumn{2}{c}{$\Delta \alpha \cos \delta$} & \multicolumn{2}{c}{$\Delta \delta$}                  \\
%
 %
  \colhead{} & \twocolhead{~($\arcsec$)~} & \twocolhead{~($\arcsec$)~} & \twocolhead{~($\arcsec$)~} & \twocolhead{~($\arcsec$)~} & \twocolhead{~($\arcsec$)~} & \twocolhead{~($\arcsec$)~} & \twocolhead{~($\arcsec$)~ } & \twocolhead{~($\arcsec$)~} & \multicolumn{2}{c}{~($\arcsec$)~} & \multicolumn{2}{c}{~($\arcsec$)~}  
%
}
\decimals
\startdata
PS J0147+4630 & $0.0000$ & $0.0000$ & $1.1697$ & $-0.4141$ & $-1.2404$ & $-0.0993$ & $-0.3394$ & $-3.2351$ & $-0.1650$ & $-2.0671$ \\
SDSS J0248+1913 & $0.0000$ & $0.0000$ & $-0.1472$ & $-0.8344$ & $0.8543$ & $-1.4463$ & $0.9050$ & $-0.0366$ & $0.4905$ & $-0.6361$ \\
ATLAS J0259-1635 & $0.0000$ & $0.0000$ & $-0.3192$ & $0.8789$ & $-1.4736$ & $0.5647 & $-0.7163$ & $-0.3845$ & $-0.6812$ & $0.2950$ \\
DES J0405-3308 & $0.0000$ & $0.0000$ & $-1.0656$ & $-0.3261$ & $-0.3416$ & $0.8332$ & $-1.2232$ & $0.6967$ & $-0.7007$ & $0.2364$ \\
DES J0420-4037 & $0.0000$ & $0.0000$ & $0.2432$ & $1.0271$ & $0.8732$ & $1.1431$ & $1.4018$ & $-0.2268$ &      $0.6938$ & $0.3468$ \\
SDSS J1251+2935 & $0.0000$ & $0.0000$ & $0.2913$ & $0.9626$ & $0.3581$ & $0.3756$ & $-1.4249$ & $0.9553$ & $-0.3459$ & $0.6239$ \\
SDSS J1433+6007 & $0.0000$ & $0.0000$ & $2.0455$ & $1.5804$ & $0.0017$ & $3.7554$ & $-0.7599$ & $1.6262$ & $0.9332$ & $1.7008$ \\
PS J1606-2333 & $0.0000$ & $0.0000$ & $-1.6211$ & $-0.5888$ & $-0.7919$ & $-0.9052$ & $-1.1278$ & $0.1504$ & $-0.8345$ & $-0.3669$ \\
DES J2038-4008 & $0.0000$ & $0.0000$ & $1.5128$ & $-0.0274$ & $1.3873$ & $2.0594$ & $-0.7923$ & $1.6769$ & $0.6779$ & $1.1833$ \\
ATLAS J2344-3056 & $0.0000$ & $0.0000$ & $-0.2915$ & $0.6696$ & $-0.8744$ & $0.3207$ & $-0.6332$ & $-0.3364$ & $-0.4173$ & $0.1515$ \\
\enddata
\end{deluxetable}

Our new astrometric measurements for the ten quadruply lensed quasars for
which \citet{luhtaru} deferentially adopted those of \citet{shajib} are presented in
Table \ref{tab:newmeas}.  The agreement between the ``provisional''  ad
hoc {\tt clumpfit} astrometry of \citet{luhtaru} and the {\tt GLEE}
  astrometry of \citet{ertl} was unexpectedly good.  If we assume that
  the two programs contribute equally to the rms differences in the
  quasar astrometry, we get errors of of $0\farcs0013$ in each
  coordinate.  We have therefore included a fourth decimal place in
  Table \ref{tab:newmeas}.

\subsection{Confirmation of shear dominance}\label{confirmation}

In Table \ref{tab:10decompositions} we give shear/ellipticity decompositions
for our 10 newly measured systems that, for the reasons elaborated above,
we believe are more reliable than those in Table 2 of \citet{luhtaru}.

\startlongtable
\begin{deluxetable}{lDDDDlDDDD}
\centerwidetable
\tabletypesize{\scriptsize}
\tablecaption{Shear ellipticity decompositions for
  10 newly measured systems} \label{tab:10decompositions}
\tablehead{
  \colhead{System name} & \multicolumn{2}{c}{$\gamma$} & \multicolumn{2}{c}{$\eta$} & \multicolumn{2}{c}{$\Delta \gamma$} & \multicolumn{2}{c}{$\Gamma_{\text eff}$} & \colhead{System name} & \multicolumn{2}{c}{$\gamma$} & \multicolumn{2}{c}{$\eta$} & \multicolumn{2}{c}{$\Delta \gamma$} & \multicolumn{2}{c}{$\Gamma_{\text eff}$} \\    
}
\decimals
\startdata
PS J0147+4630    &  0.180 & -0.020 & 0.017 & 0.160 & SDSS J1251+2935  &  0.053 &  0.053 & 0.005 & 0.105 \\
SDSS J0248+1913  &  0.120 & -0.020 & 0.012 & 0.100 & SDSS J1433+6007  &  0.197 &  0.001 & 0.038 & 0.198 \\
ATLAS J0259-1635 &  0.031 &  0.038 & 0.062 & 0.069 & PS J1606-2333    &  0.234 & -0.026 & 0.139 & 0.209 \\
DES J0405-3308   & -0.011 &  0.034 & 0.051 & 0.023 & DES J2038-4008   &  0.035 &  0.056 & 0.007 & 0.091 \\
DES J0420-4037   &  0.000 &  0.034 & 0.004 & 0.034 & ATLAS J2344-3056 & -0.168 &  0.232 & 0.016 & 0.067 \\
\enddata
\end{deluxetable}

While individual entries differ, replacement of the corresponding
entries in Table 2 of \citet{luhtaru} with the values in Table
\ref{tab:10decompositions} gives no substantial change in the
conclusion that shear dominates ellipticity by roughly a factor of
2:1.  This confirmation of the Luhtaru et al.\ result would
have been the conclusion of the paper that we originally intended to write,
but it seemed less consequential than the proximity of the lensing galaxies to Witt's hyperbola 
that we address  in the main body of this paper.



\bibliography{Schechter}{}
\bibliographystyle{aasjournal}



\end{document}